\documentclass[a4paper,11pt]{article}
\usepackage{pos}

\def\beq{\begin{equation}}
\def\eeq{\end{equation}}
\def\beqa{\begin{eqnarray}}
\def\eeqa{\end{eqnarray}}

\title{Top quark differential distributions in the SMEFT}

\author*{Nikolaos Kidonakis}
\author{Kaan \c{S}im\c{s}ek}

\affiliation{Department of Physics, Kennesaw State University,\\
Kennesaw, GA 30144, USA}

\emailAdd{nkidonak@kennesaw.edu}

\abstract{We discuss top quark processes, in particular $tW$ production and top-antitop pair production, as probes of dimension-6 Standard Model Effective Field Theory (SMEFT). We calculate top-quark double differential distributions in transverse momentum and rapidity, and we perform SMEFT fits at LO, NLO and, by adding second-order soft-gluon corrections, at approximate NNLO.}

\FullConference{43rd International Conference on High Energy Physics (ICHEP 2026)\\
30 July to 5 August , 2026\\
Natal, Brazil\\}

\begin{document}
\maketitle

\section{Introduction}

We consider $tW$ production \cite{NKKS1} and $t{\bar t}$ production \cite{NKKS2} as probes of dimension-6 SMEFT. The SMEFT Lagrangian is given by
\beq
{\cal L} = {\cal L}_{\rm SM} + \sum_{n>4} {1 \over \Lambda^{n-4}} \sum_k C_k^{(n)} O_k^{(n)}
\eeq
where  $O_k^{(n)}$ are operators of mass dimension $n>4$ \cite{smeft1,smeft2}; here, we consider $n=6$.

We parametrize the cross section at all perturbative orders as
\beq
\sigma = \sigma_{\rm SM} + \sum_k C_k \sigma_k + \sum_{k\leq k'} C_k C_{k'} \sigma_{kk'}
\eeq
with linear and quadratic SMEFT corrections. We calculate complete QCD corrections at NLO and add soft-gluon corrections at NNLO. Thus, we produce approximate NNLO (aNNLO) predictions for double-differential distributions in $p_T$ and rapidity, and we use these results to set bounds on Wilson coefficients and corresponding effective scales. 

For $tW$ production, we derive marginalized 95\% CL bounds on the dimension-six Wilson coefficients $C_{tG}$, $C_{tW}$, and $C_P$, tightening global-fit constraints by up to 14\% in the $(C_{tG}, C_{tW})$ subspace \cite{NKKS1}. For $t {\bar t}$ production, we derive marginalized bounds on the dimension-six operators $C_{tG}$, $C_u^+$, $C_d^+$, and $C_b^+$ at 13 and 13.6 TeV, improving existing global fits by up to 60\% and probing scales up to 3.9 TeV for $C_{tG}$ \cite{NKKS2}.

 \section{Higher-order soft-gluon corrections}

Soft-gluon corrections \cite{NKloop} are important for top-quark processes, both in the Standard Model \cite{NKtt} and beyond, e.g. for top-quark anomalous couplings \cite{topanom} and in SMEFT \cite{NKAT,NKKS1,NKKS2}. For the partonic processes $a(p_a)\, + \, b\, (p_b) \to t(p_t)\, + \, X $, with kinematical variables $s=(p_a+p_b)^2$, $t=(p_a-p_t)^2$, $u=(p_b-p_t)^2$, the variable $s_4=s+t+u-m_t^2-m_X^2$ goes to zero at partonic threshold.

The soft-gluon corrections appear as plus distributions of $\ln^k(s_4/m_t^2)/s_4$.
We resum these soft corrections in one-particle-inclusive (1PI) kinematics, and we provide finite-order expansions of the resummed cross section to avoid the use of prescriptions. We add the second-order soft-gluon corrections to exact NLO results and, thus, derive aNNLO predictions for cross sections and distributions.

Using Laplace transforms, we factorize the cross section into functions that describe universal soft and collinear emission from the incoming partons, a hard function $H_{ab \to tX}$ for the hard scattering, and a soft function $S_{ab \to tX}$ for noncollinear soft-gluon emission. The renormalization group evolution of these functions leads to resummation:
\beqa
{\hat{\sigma}}_{ab \to tX}^{res}(N) &=&   
\exp\left[E_a+E_b\right] \; 
{\rm tr} \left \{H_{ab \to tX} \left(\alpha_s(\sqrt{s})\right) \;
\exp \left[\int_{\sqrt{s}}^{{\sqrt{s}}/{\tilde N}} \frac{d\mu}{\mu} \;
\Gamma^{\dagger}_{S \, ab \to tX} \left(\alpha_s(\mu)\right)\right] \right.
\nonumber \\ && \hspace{-8mm} \times \, \left.
S_{ab \to tX} \left(\alpha_s(\sqrt{s}/{\tilde N}) \right) \;
\exp \left[\int_{\sqrt{s}}^{{\sqrt{s}}/{\tilde N}} 
\frac{d\mu}{\mu}\; \Gamma_{S \, ab \to tX}
\left(\alpha_s(\mu)\right)\right] \right\} \, ,
\eeqa
where $E_a$, $E_b$ are universal exponents, and the soft anomalous dimensions $\Gamma_{S \, ab \to tX}$ are known at two loops or higher.

\section{$tW$ production in SMEFT}

We begin with $tW$ production via the partonic process  $b g \to tW^-$. The relevant Feynman rules for $ttg$ and $tbW$ vertices are written in terms of three coefficients $C_{tG}$, $C_{tW}$, $C_P$ \cite{NKKS1}. We calculate top-quark double-differential distributions in transverse momentum and rapidity with MSHT20 pdf \cite{MSHT20} using bins motivated by \cite{CMS1,ATLAS}. We use the software based on \cite{sw1,sw2,sw3,sw4,sw5} for the NLO results, and we add to them the aNNLO corrections; see Ref. \cite{NKKS1} for details.

\begin{figure}[h!]
\begin{center}
\includegraphics[width=45mm]{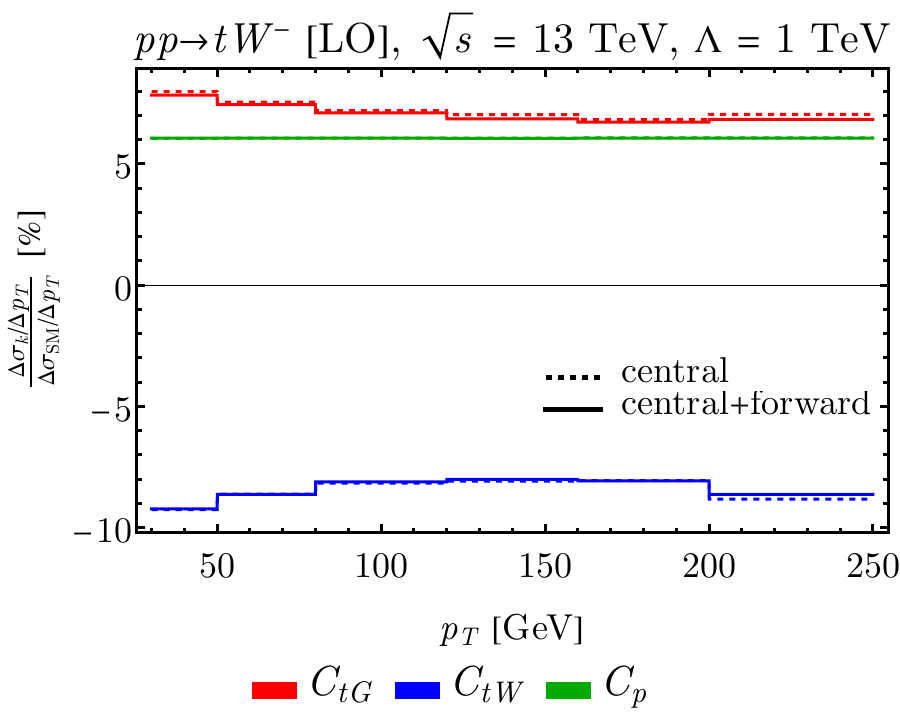}
\hspace{5mm}
\includegraphics[width=45mm]{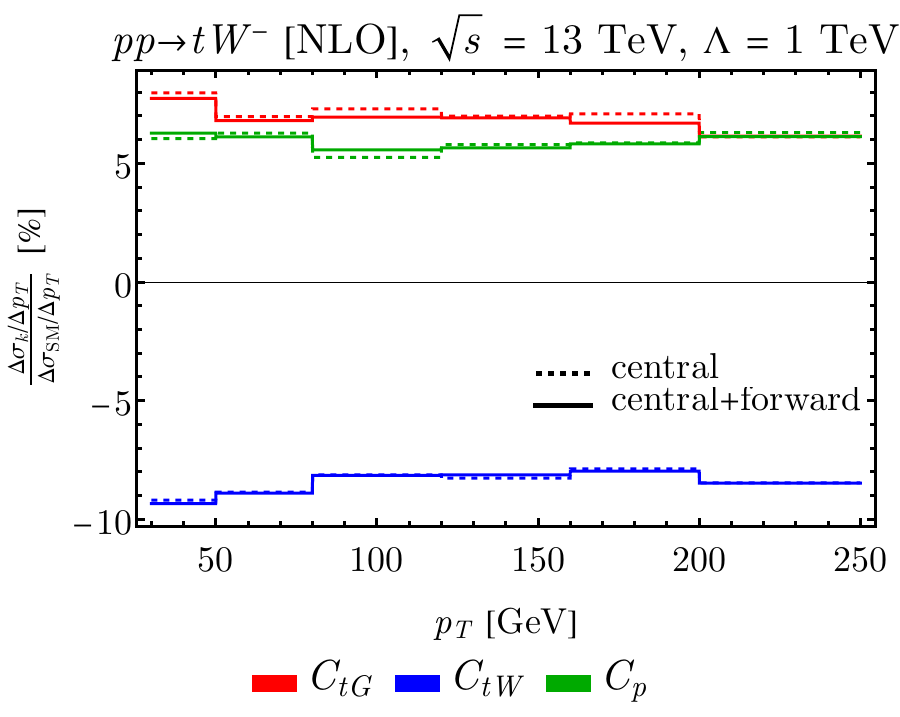}
\hspace{5mm}
\includegraphics[width=45mm]{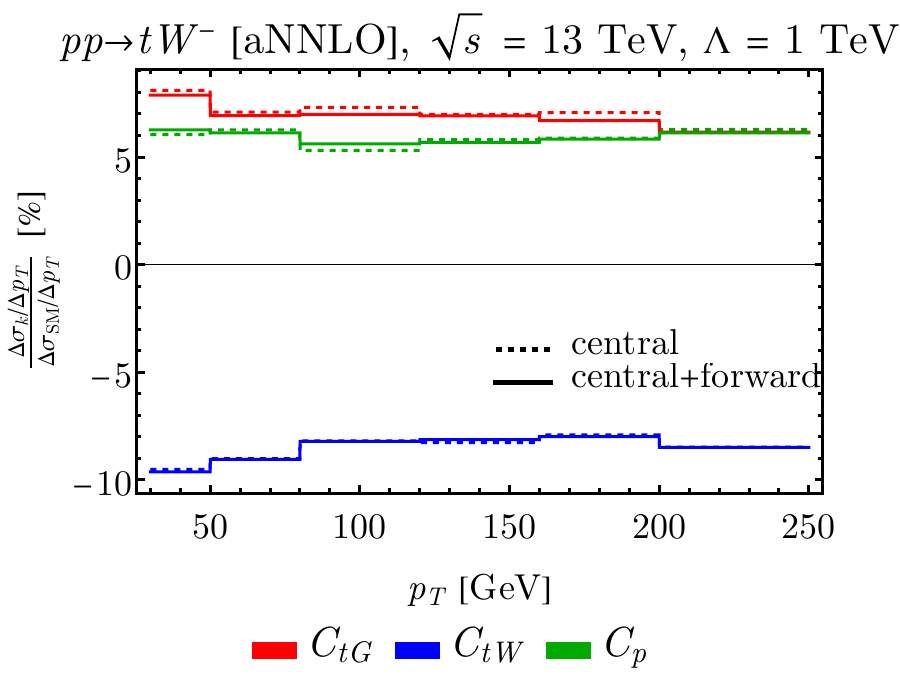}
\vspace{2mm}
\includegraphics[width=45mm]{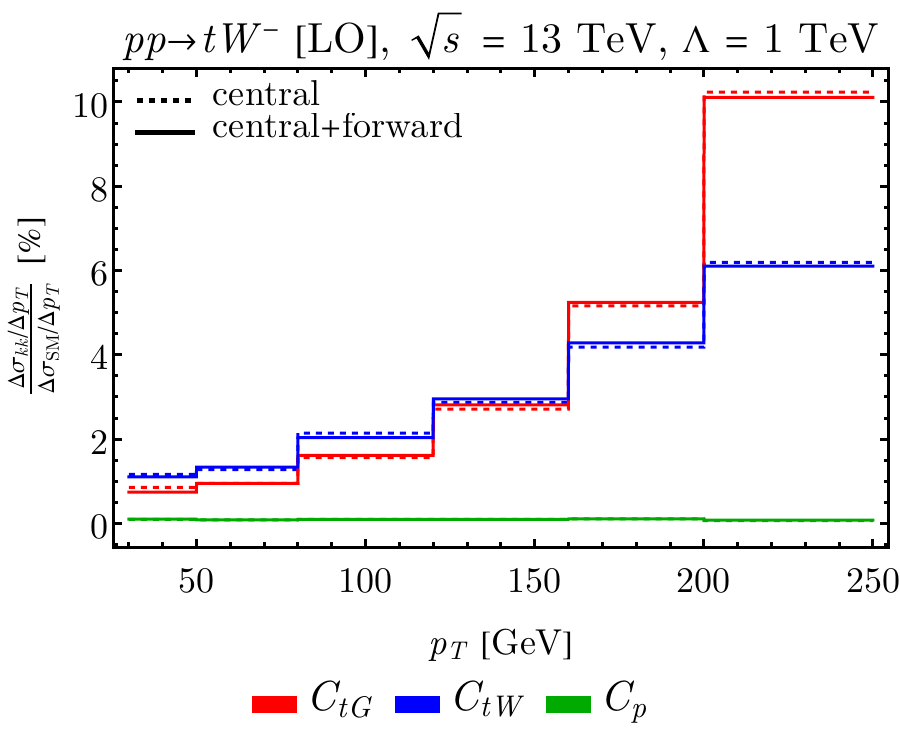}
\hspace{5mm}
\includegraphics[width=45mm]{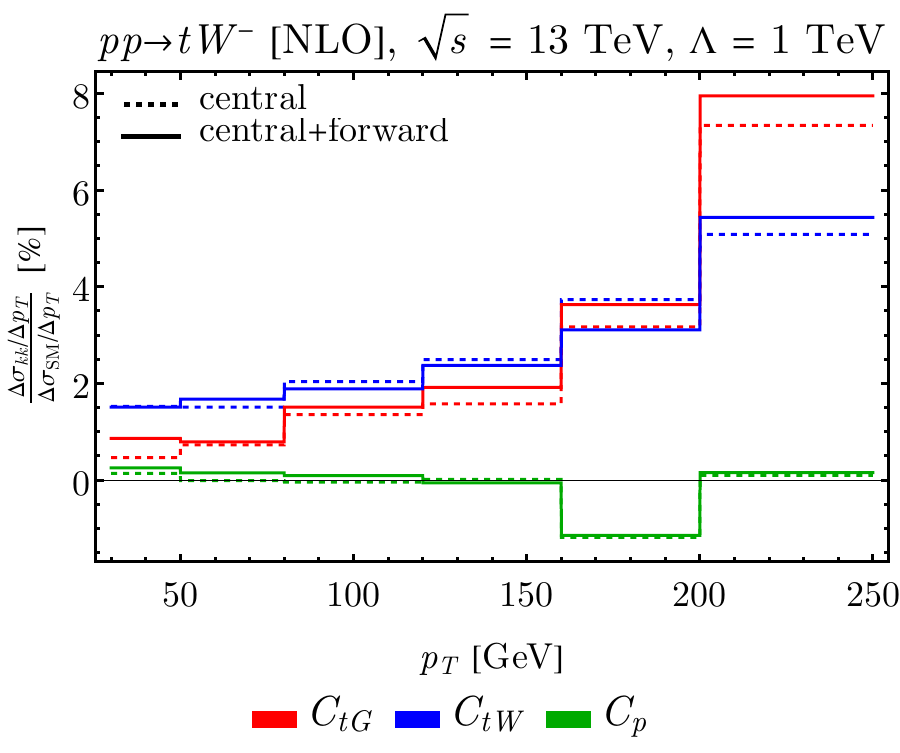}
\hspace{5mm}
\includegraphics[width=45mm]{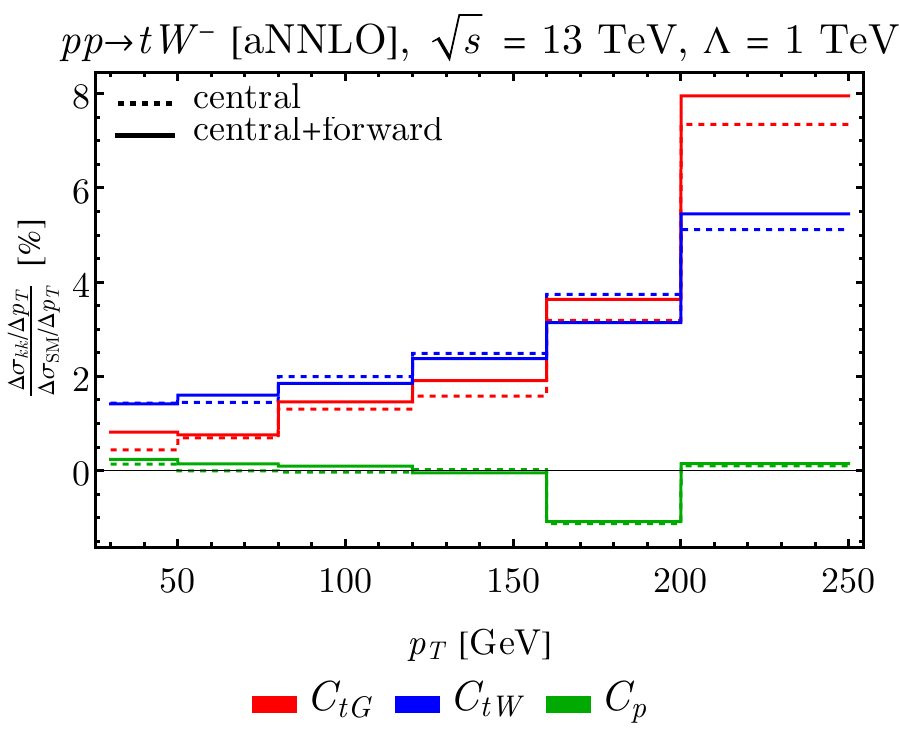}
\caption{Linear (upper plots) and quadratic (lower plots) SMEFT corrections for $C_{tG}$, $C_{tW}$, and $C_P$ relative to SM at LO, NLO, and aNNLO.}
\label{tWlinquad}
\end{center}
\end{figure}

\begin{figure}[h!]
\begin{center}
\includegraphics[width=65mm]{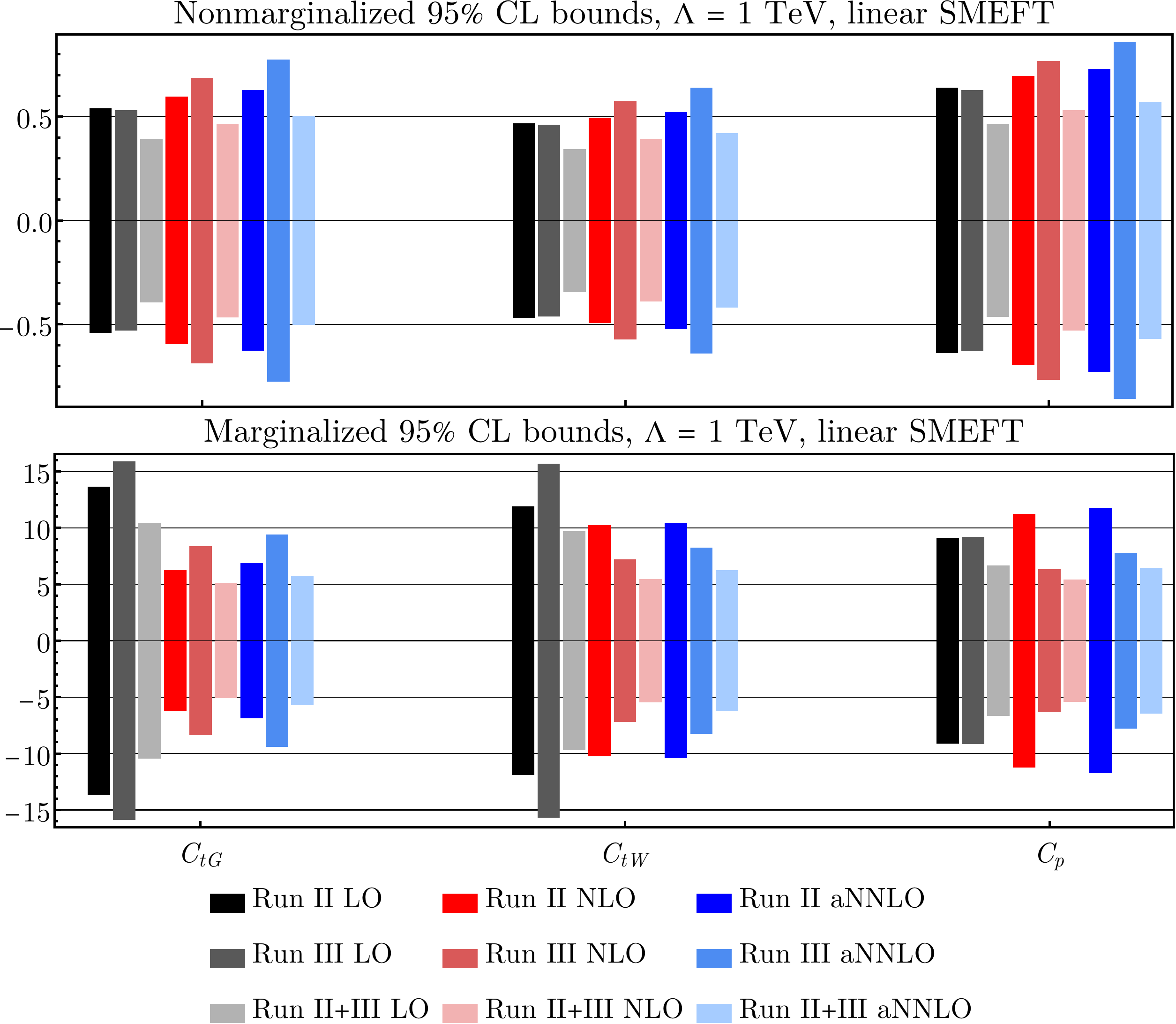}
\hspace{10mm}
\includegraphics[width=65mm]{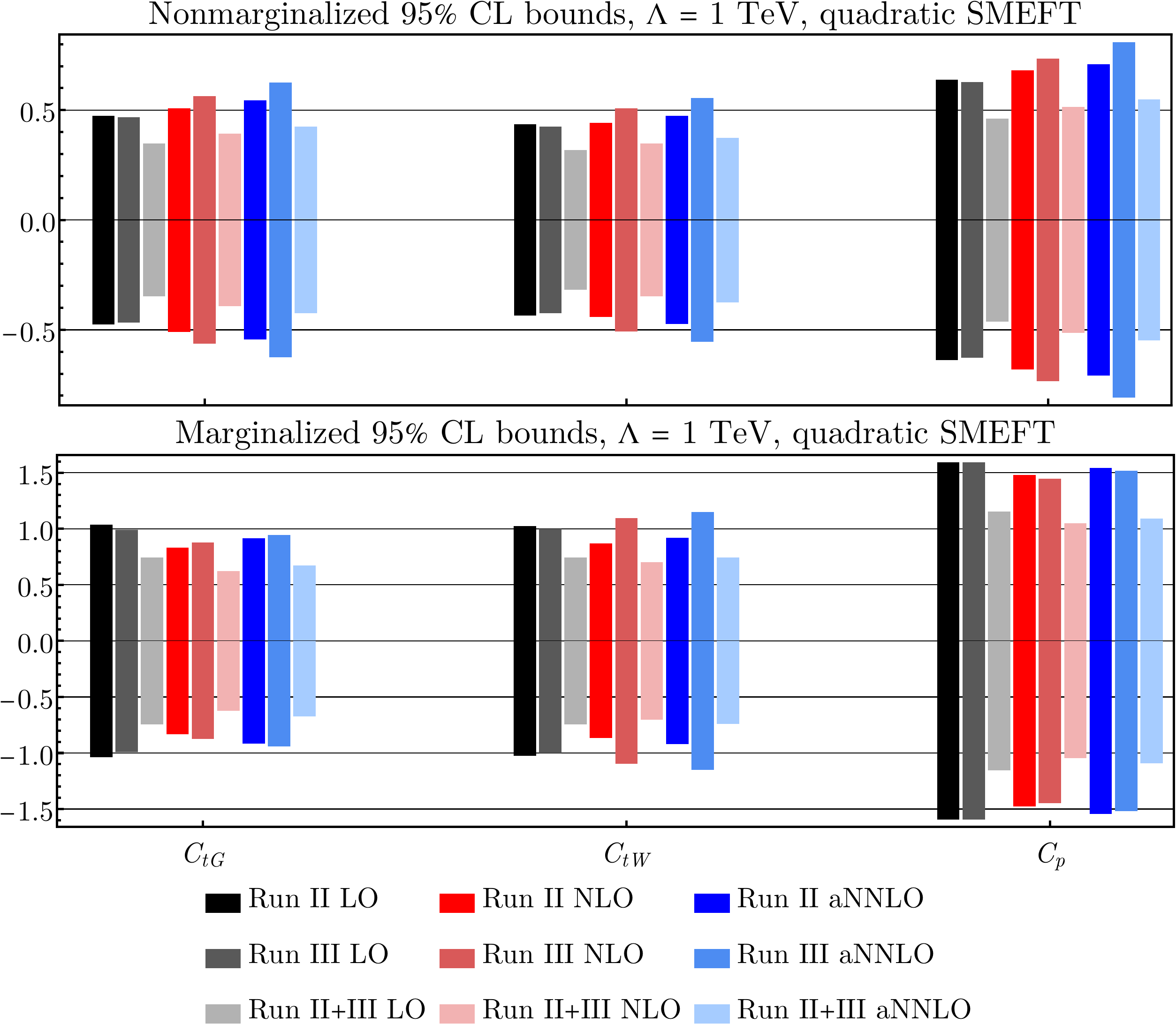}
\caption{95\% CL bounds for the linear (left) and quadratic (right) nonmarginalized and marginalized SMEFT fits.}
\label{tWfits}
\end{center}
\end{figure}

\begin{figure}[h!]
\begin{center}
\includegraphics[width=65mm]{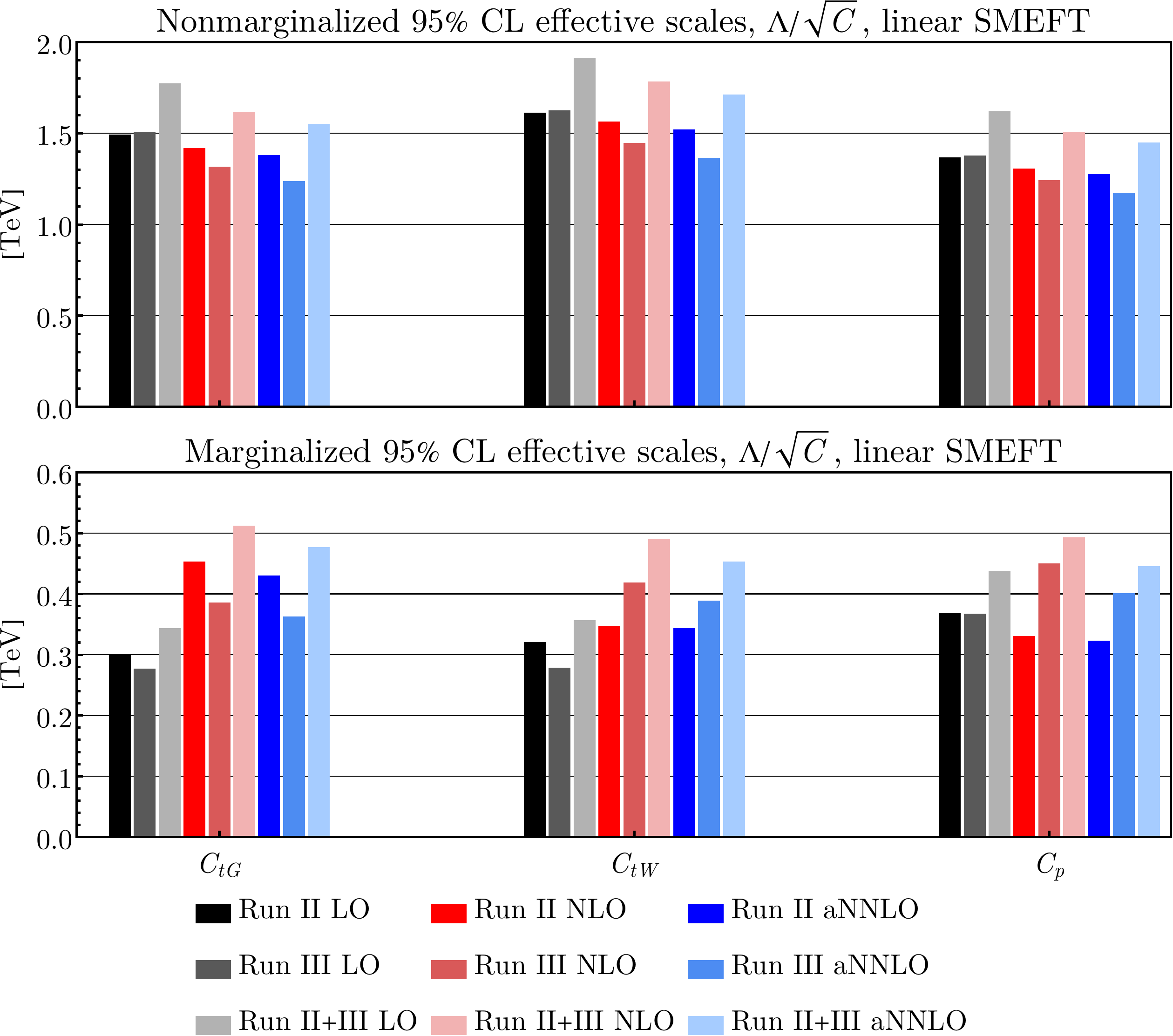}
\hspace{10mm}
\includegraphics[width=65mm]{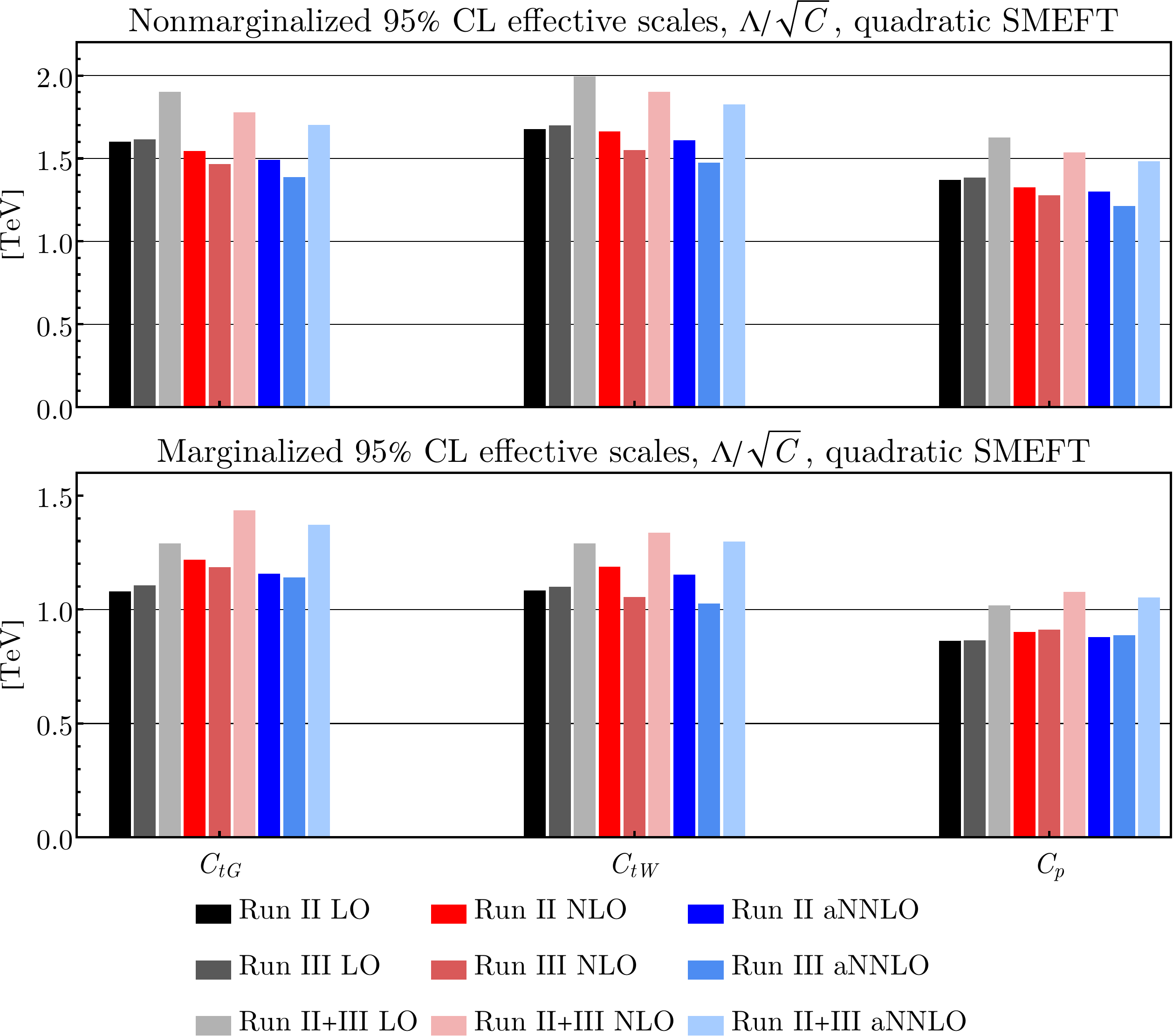}
\caption{95\% CL effective scales for the linear (left) and quadratic (right) nonmarginalized and marginalized SMEFT fits.}
\label{tWeffscales}
\end{center}
\end{figure}

In Fig. \ref{tWlinquad} we plot the linear and quadratic SMEFT corrections for the top quark $p_T$ distribution in the central region $0 < |y| < 1$, and in the forward region $1 < |y| < 3$. Fig. \ref{tWfits} shows the 95\% CL bounds for the linear and quadratic nonmarginalized and marginalized SMEFT fits
for $tW$ production. Fig. \ref{tWeffscales} shows the 95\% CL effective scales for the linear and quadratic  nonmarginalized and marginalized SMEFT fits for $tW$ production. We find that effective scales up to 2 TeV can be probed in nonmarginalized fits.

\section{$t{\bar t}$ production in SMEFT}

We continue with $t{\bar t}$ production for which the leading-order partonic processes are $q {\bar q} \to t {\bar t}$ and $ g g \to t {\bar t}$. The SMEFT Feynman rules for $ttg$, $ttgg$, $ttqq$  vertices are written in terms of four coefficients $C_{tG}$, $C_{u^+}$, $C_{d^+}$, $C_{b^+}$ \cite{NKKS2}. We calculate complete QCD corrections at NLO using \cite{sw1,sw2,sw3,sw4,sw5}, and we add soft-gluon corrections at NNLO for the top quark single- and double-differential distributions in transverse momentum and rapidity using bins as in \cite{CMS2,CMS3}. Details are provided in Ref. \cite{NKKS2}.

\begin{figure}[h!]
\begin{center}
\includegraphics[width=45mm]{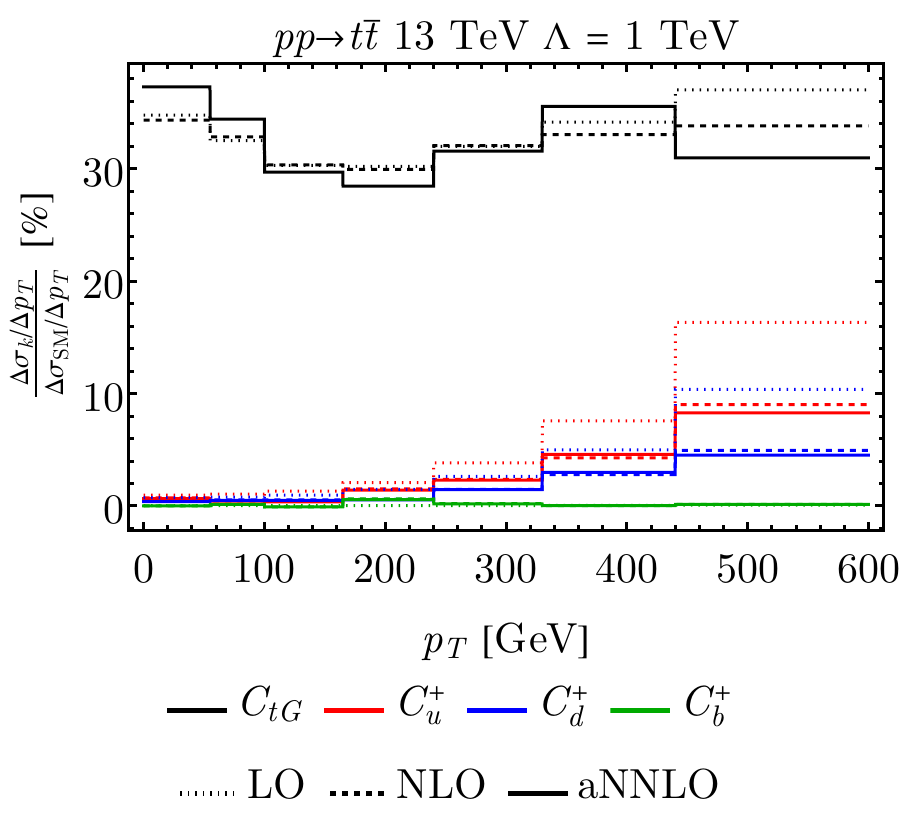}
\hspace{15mm}
\includegraphics[width=45mm]{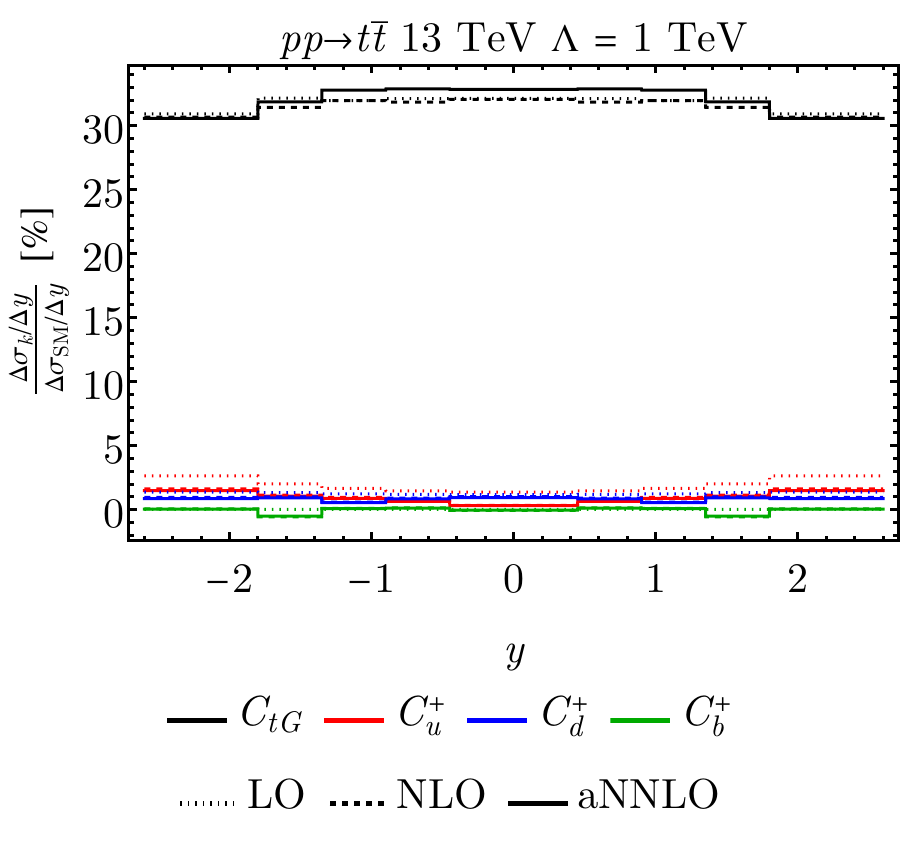}
\vspace{2mm}
\includegraphics[width=45mm]{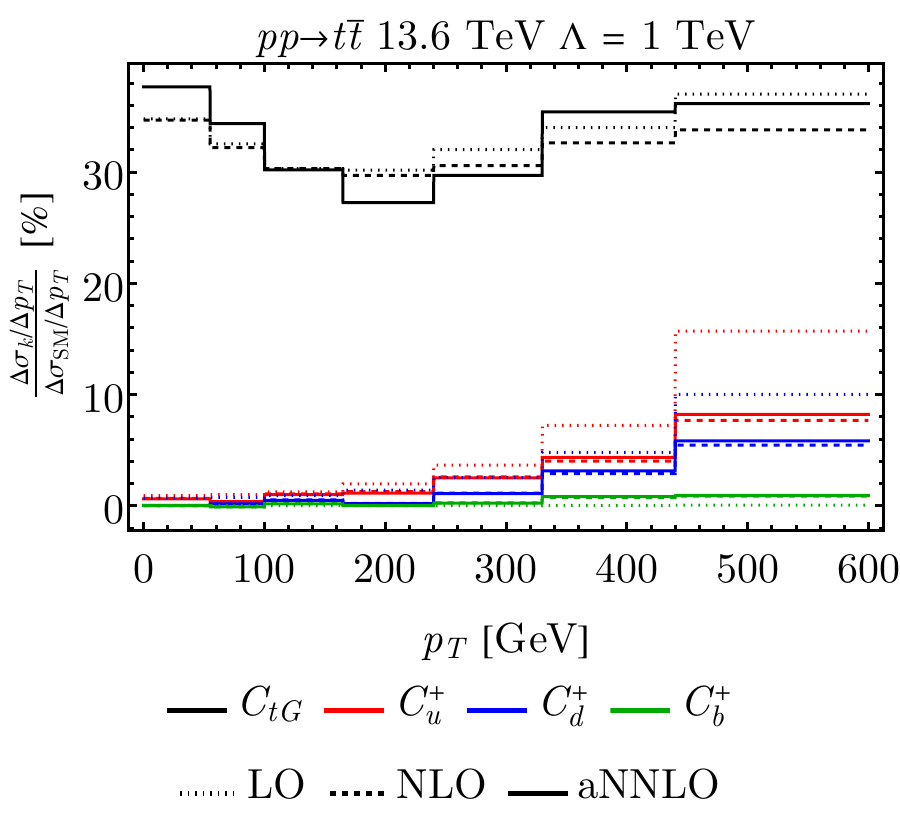}
\hspace{15mm}
\includegraphics[width=45mm]{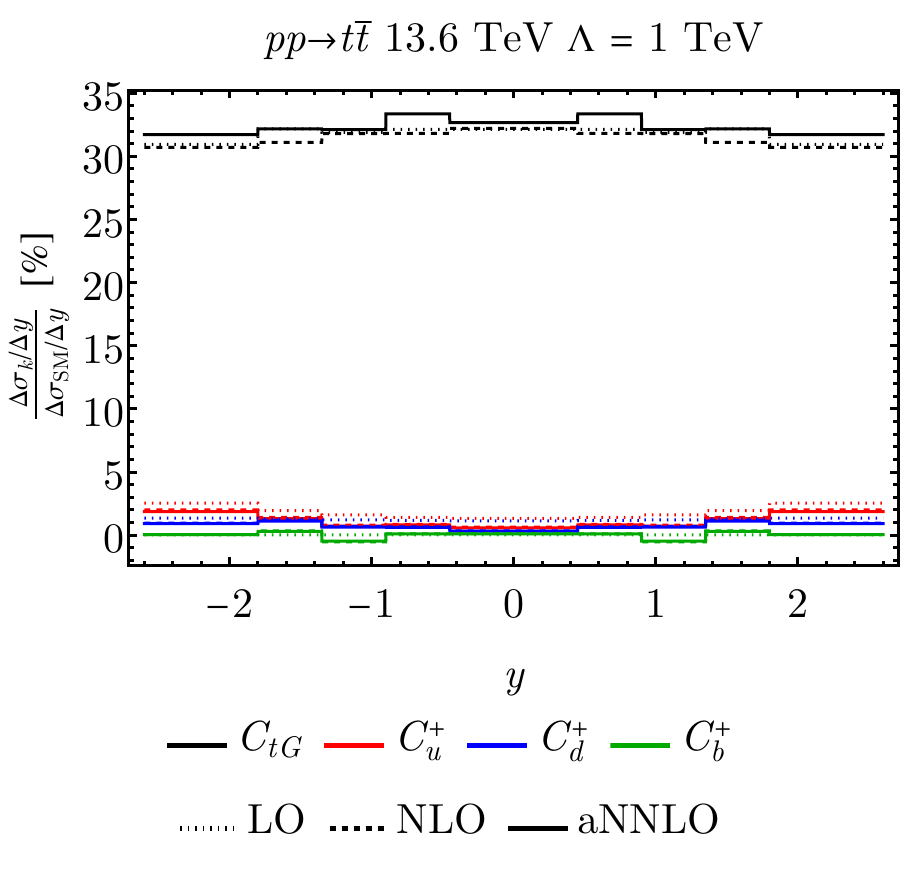}
\caption{SMEFT corrections for $C_{tG}$, $C_{u^+}$, $C_{d^+}$, $C_{b^+}$ relative to SM at LO, NLO, and aNNLO at 13 TeV (upper plots) and 13.6 TeV (lower plots).}
\label{ttpty}
\end{center}
\end{figure}

The SMEFT top-quark single-differential distributions in transverse momentum and rapidity are shown in Fig. \ref{ttpty} at 13 TeV (upper plots) and 13.6 TeV (lower plots).

\begin{table}[h!]
\begin{center}
\begin{tabular}{|c|c|c|c|c|} \hline 
 & $C_{tG}$ & $C_u^+$ & $C_d^+$ & $C_b^+$  \\ \hline 
   LO & 0.16 & 0.74 & 1.3 & $1.4\times 10^2$ \\
   NLO & 0.083 & 0.57 & 0.98 & 4.3 \\
   aNNLO & 0.059 & 0.48 & 0.81 & 2.8 \\  \hline 
 & ${\Lambda/\sqrt{\Delta C_{tG}}}$ & ${\Lambda/\sqrt{\Delta C_u^+}}$ & ${\Lambda/\sqrt{\Delta C_d^+}}$ & ${\Lambda/\sqrt{\Delta C_b^+}}$ \\ \hline 
   LO & 2.5 & 1.2 & 0.89 & 0.085 \\ 
   NLO & 3.5 & 1.3 & 1.0 & 0.48 \\
   aNNLO & 4.1 & 1.4 & 1.1 & 0.60 \\ \hline 
\end{tabular}
\caption{95\% CL non-marginalized bounds on $C_{tG}$, $C_u^+$, $C_d^+$, and $C_b^+$, and effective scales.}
\end{center}
\end{table}

\begin{table}[h!]
\begin{center}
\begin{tabular}{|c|c|c|c|c|} \hline 
 & $C_{tG}$ & $C_u^+$ & $C_d^+$ & $C_b^+$  \\ \hline 
   LO & 0.55 & 5.0 & 13 & $9.1\times 10^2$ \\  
   NLO & 0.10 & 1.6 & 2.8 & 5.0 \\ 
   aNNLO & 0.066 & 1.1 & 1.9 & 3.1 \\  \hline 
 & ${\Lambda/\sqrt{\Delta C_{tG}}}$ & ${\Lambda/\sqrt{\Delta C_u^+}}$ & ${\Lambda/\sqrt{\Delta C_d^+}}$ & ${\Lambda/\sqrt{\Delta C_b^+}}$ \\  \hline 
   LO & 1.3 & 0.45 & 0.28 & 0.033 \\ 
   NLO & 3.2 & 0.80 & 0.60 & 0.45 \\ 
   aNNLO & 3.9 & 0.95 & 0.72 & 0.57 \\ \hline 
\end{tabular}
\caption[]{95\% CL marginalized bounds on $C_{tG}$, $C_u^+$, $C_d^+$, and $C_b^+$, and effective scales.}
\end{center}
\end{table}

Tables 1 and 2 display, respectively, the non-marginalized and marginalized 95\% CL bounds on Wilson coefficients $C_{tG}$, $C_u^+$, $C_d^+$, and $C_b^+$ together with the corresponding effective scales $\Lambda/\sqrt{\Delta C}$. The higher-order corrections give significantly more robust bounds, 
and effective scales up to around 4 TeV can be probed.

\acknowledgments

This material is based upon work supported by the National Science Foundation under Grant No. PHY 2412071.

\end{document}